\documentclass[a4paper]{article}
\usepackage{placeins}
\usepackage{pbox}
\usepackage{graphicx}
\usepackage{amsmath}
\usepackage{amssymb}
\usepackage{MathShorthand}
\usepackage{subfigure}
\usepackage{verbatim}
\usepackage{epstopdf}
\usepackage[affil-it]{authblk}
\usepackage{float}
\usepackage{amsfonts}%
\usepackage[margin=0.75in]{geometry}
\providecommand{\U}[1]{\protect\rule{.1in}{.1in}}

\begin{document}
\title{Drunken robber, tipsy cop: First passage times, mobile traps, and Hopf bifurcations}
\author{Justin C. Tzou$^1$, Shuangquan Xie$^1$ and Theodore
Kolokolnikov$^1$}
\affil{$^1$Department of Mathematics and Statistics, Dalhousie University,
Halifax, Canada}
\date{\today }
\maketitle

\begin{abstract}
For a random walk on a confined one-dimensional domain, we consider mean first
passage times (MFPT) in the presence of a mobile trap. The question we address is
whether a mobile trap can improve capture times over a stationary trap. We
consider two scenarios: a randomly moving trap and an oscillating trap. In
both cases, we find that a stationary trap actually performs better (in terms of reducing expected capture time) than a very
slowly moving trap; however, a trap moving sufficiently fast performs better than a
stationary trap. We explicitly compute the thresholds that separate the two
regimes. In addition, we find a surprising relation between the oscillating
trap problem and a moving-sink problem that describes reduced dynamics of a
single spike in a certain regime of the Gray-Scott model. Namely, the
above-mentioned threshold corresponds precisely to a Hopf bifurcation that
induces oscillatory motion in the location of the spike. We use this correspondence
to prove the uniqueness of the Hopf bifurcation.
\end{abstract}

\baselineskip=12pt

\vspace{16pt}

\noI \textbf{Key words: mean first passage time, mobile trap, Hopf-bifurcation, reaction-diffusion, matched asymptotics, boundary layer}

\vspace{16pt}

\setcounter{equation}{0}

\section{Introduction}

Numerous problems in nature can be formulated in terms of expected escape time of
Brownian particles in the presence of traps. This escape time is often referred to as the mean first
passage time (MFPT). For example, a cell is regulated by chemical reactions
involving a small number of signaling molecules that have to find their
targets in a complex and crowded environment \cite{schuss2007narrow}. Further
examples include oxygen transport in muscle tissue \cite{titcombe2000asymptotic},
cold atoms in optical traps \cite{barkai2013transport}, molecular
self-assembly \cite{yvinec2012first}, optimal search strategies \cite{benichou2005optimal, benichou2011intermittent}, and proteins searching for target sequences on a DNA strand \cite{lomholt2005optimal, sokolov2005target, ramanathan2009type, mirny2009protein, benichou2009searching, benichou2014first}.

In a recent review of MFPT processes on confined domains \cite{benichou2014first}, it was remarked that while the case of stationary traps is well studied, MFPT problems with mobile traps in confined domains still remain largely unexplored. It is only recently
that attention has shifted to mobile traps \cite{giuggioli2013encounter,
holcman2009probability, tejedor2011residual, tzou2014mean}, which is not only
more realistic in many situations, but can significantly alter the system's
behaviour. In this paper, we formulate two MFPT problems with a mobile trap on a confined one-dimensional domain
and study the effect of trap motion on the average MFPT.

Mobile traps occur naturally in a variety of scenarios. An early example was introduced in \cite{toussaint1983particle} for the annihilation reaction $A + B \to 0$, motivated by the annihilation of mobile monopole-antimonopole pairs in the early universe. This general model may also describe chemical kinetics in diffusion-limited regimes \cite{rice1985diffusion,hanggi1990reaction} where reaction rates are limited by the encounter rate of the reactants, and collision-induced quenching of excited-state particles \cite{szabo1988diffusion}. Other examples that fit into the mobile traps framework include ligands binding to a receptor on a non-stationary cell, and disease spread \cite{giuggioli2013encounter}, where susceptible walkers become infected upon encounter with infected walkers. Perhaps the most common example is that in which predators search for mobile prey \cite{komarov2013capturing, oshanin2009survival, oshanin2002trapping, gabel2012can, kehagias2013cops}. Related MFPT problems with stationary traps involve random walks in the presence of time-constant \cite{kurellaasymptotic,bressloff2013stochastic} and time-fluctuating fields \cite{fletcher1988first, JavierBrey1994123, revelli2004diffusion, dhara2002coherent, dybiec2004resonant, pikovsky1997coherence}.

The following scenario illustrates the types of questions that the
prototypical MFPT problems below seek to address. Consider a child in a
mall that becomes separated from her father. Unable to remember where she
initially became separated, the girl performs a random walk in an attempt
to locate her father. The question then becomes what the father 
should do in order to find his daughter in the shortest amount of time.
While he might instinctively move about in an active search for the
child, it may in fact be more beneficial to remain stationary. The answer
depends on the father's initial location relative to physical boundaries,
how fast he moves in relation to the daughter, and what type of motion he
follows. 

On an infinite domain, it was claimed in \cite{blythe2003survival} for continuous space and proven in \cite{moreau2003pascal, moreau2004lattice} for a discrete lattice, that a mobile target in the presence of randomly distributed Brownian traps is always expected to be captured more quickly than a stationary target. In our prototypical examples below, we show that finite domain effects can cause motion to delay expected capture time when the motion is too slow. We note that, unlike \cite{blythe2003survival} and \cite{moreau2003pascal}, we reverse the role of target and traps so that we compare average capture times in the presence of a stationary versus mobile trap.

We now state the MFPT problems that we analyze below. We consider a particle undergoing a random walk inside an isolated one-dimensional
interval, while the trap moves according to the following
mechanisms: (A) it undergoes a time-oscillatory motion about the center of
the domain with a prescribed frequency and amplitude; and (B) a random walk with prescribed diffusion rate. In each problem, we formulate an associated boundary value problem for the MFPT and compute asymptotic solutions to calculate critical trap speeds below which a mobile trap leads to longer capture times. In a manner analogous to that employed in \cite{giuggioli2013encounter, gabel2012can, redner2001guide}, we show below that the MFPT associated with these two scenarios may be obtained by solving the following two systems:
\newline

\noI \textbf{Problem A}: Randomly moving particle with an oscillating trap and
reflective endpoints:\bes\label{pbmA}%
\BE
u_{xx}+u_{t}  =-1 \,; \qquad u_{x}\left(0,t\right)=0=u_{x}\left(1,t\right) \,, \quad u(x,0)=u(x,2\pi/\omega) \,;
\EE
\BE \label{pbmAbc}
u(\xi(t),t) = 0 \,, \quad \xi(t)=\frac{1}{2}+\varepsilon\sin\left(\omega t\right) \,.
\EE
\ees 

\noI This problem corresponds to a backwards heat equation. Here, $u(x,t)$
denotes the dimensionless mean first passage time (i.e. the expected time to reach the
trap)\ for a particle located at space location $x\in\left(  0,1\right)$ at
time $t$ within the period of the trap. The trap is assumed to oscillate around the center $x=1/2$ with
frequency $\omega$ and amplitude $\varepsilon$, while the endpoints at $x = 0, 1$ are reflective. For a random walker whose initial location and start time are uniformly distributed over $x\in(0,1)$ and $t\in(0,2\pi/\omega)$, respectively, the expected MFPT is given by

\begin{equation} \label{ubar}
\bar{u}=\frac{\omega}{2\pi}\int_{0}^{\frac{2\pi}{\omega}}\int_{0}^{1} \!u(x,t)\, dxdt \,.
\end{equation}

\noI Surprisingly, this problem is also intimately connected to
oscillatory spike motion in certain reaction-diffusion systems such as
the Gray-Scott model \cite{chen2009oscillatory, 
muratov2002stability,
muratov2000static,
doelman1997pattern,
kolokolinikov2006slow}. This connection will be studied below in Problem C.
\newline

\noI \textbf{Problem B:}\ Randomly moving particle and trap:\bes\label{pbmB}%
\BE
u_{xx}+a^{2}u_{yy}=-1 \,, \qquad \left(x,y\right) \in \left(0,1\right)^{2} \,;
\EE
\BE \label{pbmBbc}
\partial_{n}u=0\text{ for }\left(  x,y\right)  \in\partial\left( \left(
0,1\right)^{2}\right) \,, \quad u =0\text{ when }x=y \,.
\EE
\ees 
\noI Here, $u\left(  x,y\right)$ is the dimensionless expected time
for a particle starting at location $x$ to hit the trap starting at location $y$. The constant $a^2$ is the ratio of trap and particle diffusivities. For a randomly moving particle whose starting location is uniformly distributed on $(0,1)$, the expected MFPT is given by

\begin{equation} \label{avgmfptrandintro}
\bar{u}(y_{0};{a})=\int_{0}^{1}\!u(x,y_{0})\,dx \,.
\end{equation}

\noI In \eqref{avgmfptrandintro}, $y_0$ is the initial location of the randomly diffusing trap. The dependence of $\bar{u}$ in \eqref{avgmfptrandintro} on $a$ is through that of $u$. We emphasize that the diffusivities of the two random walkers may be different (i.e, $a \neq 1$). For $a = 1$, an exact solution of \eqref{pbmB} may be sought using the method of images. This approach was taken in \cite{tejedor2011encounter}, though instead of solving \eqref{pbmB} directly, the MFPT was obtained from the solution of the diffusion equation on the same domain with same boundary conditions.

Finally, we show that Problem A is intimately connected to the following
problem arising in reaction-diffusion models:
\newline

\noI \textbf{Problem C:} A moving sink problem arising from reduced dynamics of
an interior spike in the Gray-Scott model:\bes\label{pbmC}%
\BE
u_{t}  =u_{xx}+1-\delta(\xi(t)-x) \,, \qquad u_{x}\left(  0,t\right)=0=u_{x}\left(  1,t\right)\,;
\EE
\BE
\xi^{\prime}(t)  =\beta\left(  u_{x}(\xi(t)^-,t)+u_{x}(\xi(t)^+,t)\right)\,. 
\EE
\ees Here, $\xi(t)$ represents the location of the spike as a function of (rescaled)\ time. In Appendix \ref{gsderiv}, we show how \eqref{pbmC} is obtained from a certain regime of the Gray-Scott model.

We now summarize our results. For problem A, in the limit of small amplitude $\varepsilon\ll1$, we calculate a critical frequency $\omega_{c}$ such that a trap with frequency $\omega>\omega_{c}$ performs more optimally (in terms of reducing the average MFPT $\bar{u}$) than a stationary trap. Conversely, for $\omega<\omega_{c}$, trap motion impairs capture time. An algebraic equation for $\omega_c$ is given in \S \ref{sec:osc}. For problem B, we show that the comparison depends both on the
initial location of the trap as well as its speed relative to that of the particle
(denoted by $a$). In particular, for a trap initially located at the center of
the domain, we find that the moving trap is more optimal when $a>a_c\approx0.53$. In both of these problems, the existence of a critical trap mobility may be inferred by a simple argument. When the trap is stationary ($\omega = a = 0$), symmetry dictates that the optimal trap location be at the center of the domain. All other locations are suboptimal. When the trap is barely mobile ($\omega, a \ll 1$), it spends almost all of its time remaining nearly stationary at suboptimal locations. The average MFPT in this case must be larger in comparison to when the trap remains stationary at its optimal location. However, in the limit of high mobility ($\omega, a \gg 1$), the trap may be thought of as everywhere at once, including at the optimal stationary location. In this case, the average MFPT would clearly be smaller in comparison to the stationary trap. There must then exist a ``critical mobility'' for which the average MFPT's associated with the stationary and mobile traps are equal.

In problem B, we also investigate how a mobile trap compares to a stationary trap not located optimally at the center of the domain. In the limit of slow trap diffusivity ($a \ll 1$), we find that a stationary trap is more optimal as long as its location $y$ is not \textquotedblleft too close\textquotedblright\ to the boundaries; that is, provided that $y\in\left(  y_{c},1-y_{c}\right)$ where $y_{c}=O(a\log a)$. The precise value of $y_c$ is given in \S \ref{brown}.

By analogy to \cite{komarov2013capturing}, we refer to problem B as the
\emph{\textquotedblleft drunken robber, drunken cop\textquotedblright}
problem: both robber and cop are drunk (at different levels of intoxication), and
the more they drink, the faster they are assumed to stagger about. Referring to \eqref{pbmB} where $x$ and $y$ denote the location of the robber and cop, respectively, the case of small $a$ that we study analytically below corresponds to a
\textquotedblleft drunk robber and tipsy cop\textquotedblright. Roughly
speaking, the conclusion is that it is more optimal for the cop to be sober than to
be slightly tipsy (as long as he is starting at an ``advantageous'' location not too close to the boundaries), but it is better still for the cop to be highly inebriated (large $a$) \footnote{This assumes, rather unrealistically, that the speed of the random walk increases with increased inebriation.}.

Finally, we show the following unexpected relationship between Problems A and C.
For small $\beta$ in \eqref{pbmC}, one can show that $\xi\rightarrow1/2$ at $t\rightarrow
\infty$. This equilibrium state becomes unstable due to a Hopf bifurcation as
$\beta$ is increased past some $\beta_{hopf}$. At the Hopf bifurcation, we show that the oscillation frequency of $\xi$ is precisely the
critical frequency $\omega_{c}$ from Problem A. Furthermore, we exploit this
connection to rigorously prove both the existence and uniqueness of
$\beta_{hopf}$. This appears to be a general phenomenon; for example, this
equivalence (as well as existence and uniqueness) still holds if $u_{xx}$ is
replaced by $u_{xx}-\mu u$ in both (\ref{pbmA}) and (\ref{pbmC}).

\setcounter{equation}{0}
\section{Oscillatory trap\label{sec:osc}}

\begin{figure}[htbp]
\begin{center}
\mbox{
\subfigure[one dimensional random walk with mobile trap] 
{\label{oneDschematic}
\includegraphics[width=.4\textwidth]{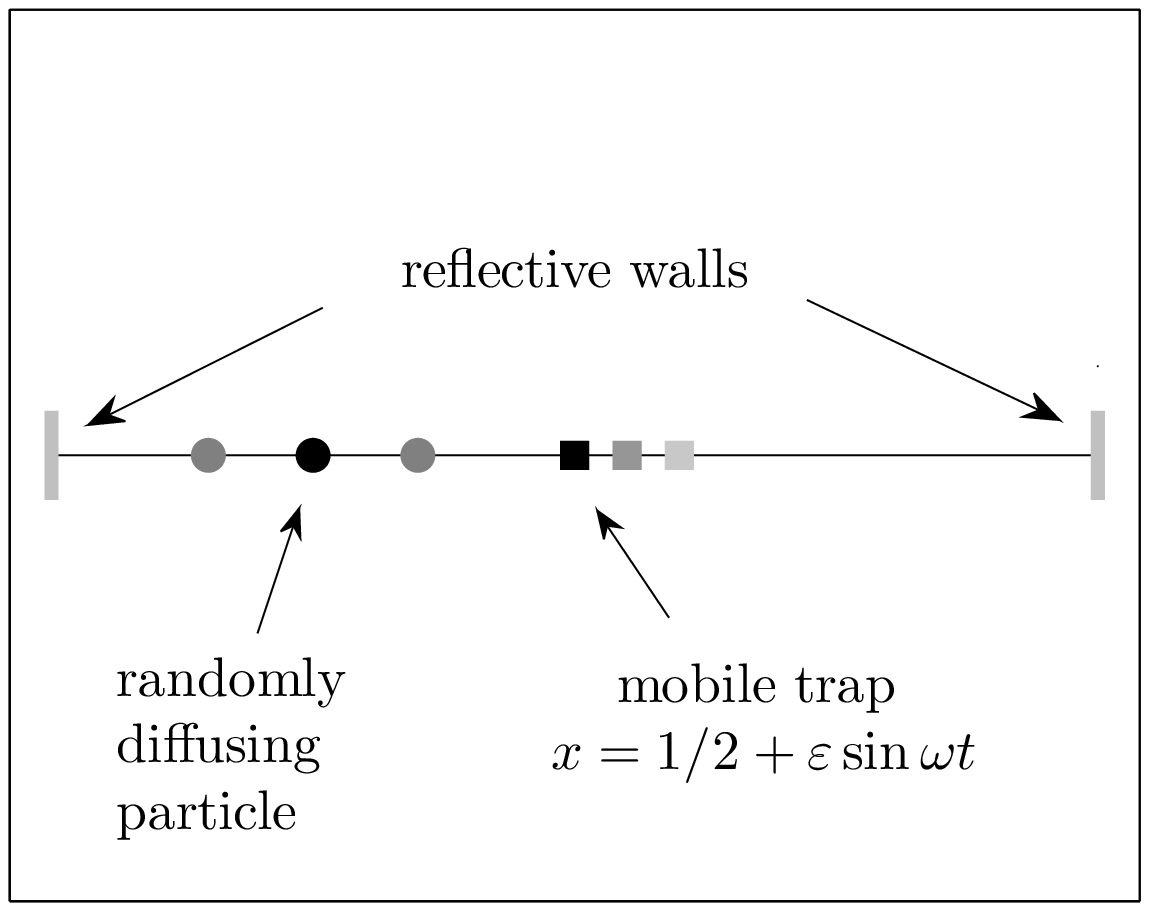}
}   \hspace{.5cm}
\subfigure[equivalent two-dimensional random walk with drift] 
{\label{twoDschematic}
\includegraphics[width=.4\textwidth]{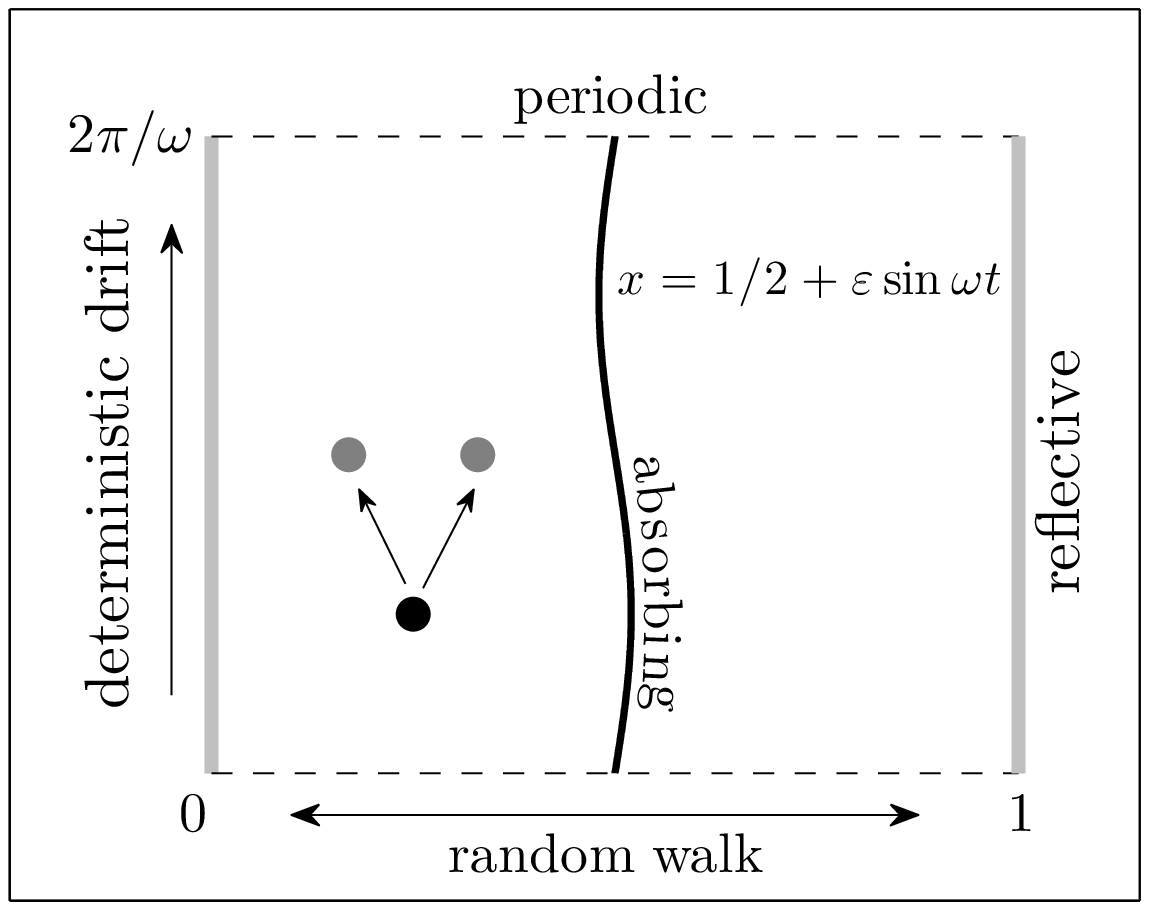}
}}
\caption{(a) A particle undergoing an unbiased random walk with a reflective left and right boundary and a mobile trap oscillating about the center of the domain. (b) The equivalent two-dimensional random walk with deterministic upwards with unit velocity. The thick gray vertical lines at $x = 0$ and $x = 1$ represent reflective walls. The domain is periodic in $t$, so a particle exiting through the top re-enters at the bottom.}
\end{center}
\end{figure}

Let us now derive (\ref{pbmA}). Consider a trap that is oscillating around the
center of the domain with a given frequency and a small amplitude. We assume
insulating boundary conditions. This situation is shown schematically in
Figure \ref{oneDschematic}. In contrast to cases with stationary traps
(see \cite{benichou2014first} and references therein), the MFPT associated with a location $x$ changes in time due to the motion of the trap. The state of this system may be defined in terms of two variables:
the location of the particle, and the location of the trap. Each time step,
the particle takes one step either to the left or to the right, while the trap
location $\xi(t)$ moves according to the deterministic function $\xi
(t)=1/2+\varepsilon\sin(\omega t)$. This suggests that the one-dimensional random walk depicted in Figure \ref{oneDschematic} may be mapped to the equivalent two-dimensional random walk shown in Figure \ref{twoDschematic}. Here, the time (vertical) axis assumes the role of the second spatial dimension. The two thick vertical gray lines at $x = 0$ and $x = 1$ represent the reflective walls, while the $2\pi/\omega$-periodicity in the vertical $t$-axis (horizontal dashed lines) is a consequence of periodic trap dynamics. The mobile trap is mapped to an interior absorbing segment indicated by the solid black curve that divides the domain into a left and right half. Since particles can only stay in one half for their entire lifetime, the mobile trap acts as a spatially dependent absorbing boundary when only the left or right half of the domain is considered (see below).

For a two-dimensional walk with random dynamics in the horizontal $x$-direction and deterministic drift in the positive $t$-direction, a static equation for the MFPT $u(x,t)$ associated with location $x$ at time $t$ may be readily derived. In analogy to the derivation of \cite{redner2001guide} for a discrete random walk with stationary traps, we write

\begin{equation} \label{mfptdiscrete}
u(x,t)=\frac{1}{2}\left\lbrack u(x+\Delta x,t+\Delta t)+u(x-\Delta x,t+\Delta t)\right\rbrack  +\Delta t\,.%
\end{equation}

\noI Equation \eqref{mfptdiscrete} states that the MFPT of a particle located at
location $(x,t)$ in the two-dimensional domain is the average of the MFPT associated with the two locations that the
particle will next occupy with equal probability, plus the $\Delta t$ time
that it takes to move there. Expanding the right-hand side of
\eqref{mfptdiscrete} for small $\Delta x$ and $\Delta t$, and defining
$D\equiv\Delta x^{2}/(2\Delta t)$, we obtain the following backward-time
diffusion equation,

\bes
\BE \label{unscaledeq}
Du_{xx}+u_{t} = -1 \,, \qquad u_{x}\left(0,t\right)=0=u_{x}(1,t) \,, \quad u(x,0) = u(x, 2\pi/\omega) \,;%
\EE
\BE
u(\xi\left(  t\right) ,t)  =0 \,, \quad \xi(t)=\frac{1}{2}+\varepsilon \sin\left(  \omega t\right) \,.
\EE
\ees 

\noI The constant $D$ in \eqref{unscaledeq} can be scaled to unity without loss of generality, which leads to the boundary value problem in \eqref{pbmA}. We remark that \eqref{mfptdiscrete} is in contrast to a regular diffusive process, for which the state at $(x,t)$ is an average of states at an earlier time $t - \Delta t$. This is due to the fact that particles captured by the interior absorbing segment in Figure \ref{twoDschematic} have left the system and thus cannot pass information in the direction of drift; instead, information from the absorbing segment propagates in the direction opposite the drift.

The solution of \eqref{pbmA} may be computed numerically by solving the
associated forward-time diffusion equation (obtained by time-reversing
$t\rightarrow-t$ in (\ref{pbmA})), which quickly converges to a $2\pi/\omega$
time-periodic solution. Alternatively, one can also solve the boundary value
problem associated with the $t$-periodic boundary conditions in
(\ref{pbmA}). We adopt the latter approach below, using FlexPDE
software\footnote{FlexPDE is a general-purpose finite element method
software, see http://www.pdesolutions.com}
to solve the associated boundary-value problem. In Figures \ref{dynslice1} and \ref{dynslice2}, we compare the MFPT as given by the PDE solution of \eqref{pbmA} with $\ve = 0.2$ and $\omega = 80$ versus that given by Monte Carlo simulations. The figures depict the MFPT associated with each location in space, at a given instant during the cycle of the trap dynamics. While the trap is located near $x = 0.5$ in both figures, the MFPT differs greatly due to the direction of motion of the trap. In particular, the MFPT is high directly behind the trap, and low directly in front of it. Figure \ref{pdesurf_osc} shows the corresponding space-time plot.

The Monte Carlo results were generated as follows. At a given time $t_m \in \lbrack0, T)$ during the cycle of trap, 10000 particles are placed at a particular point in space $x_n$. Each particle undergoes a random walk and the time to capture is recorded for each. The average of the capture times is then recorded as the MFPT associated with location $x_n$ at time $t_m$ in the trap's cycle. Repeating over each location of the discretized domain, we generate a figure of the type in Figures \ref{dynslice1} and \ref{dynslice2}. The procedure is then repeated for multiple values of $t_m$ to capture the time-dependence.

\begin{empty}\begin{figure}[htbp]
  \begin{center}
    \mbox{
    \subfigure[trap moving leftward] 
        {\label{dynslice1}
        \includegraphics[width=.3\textwidth]{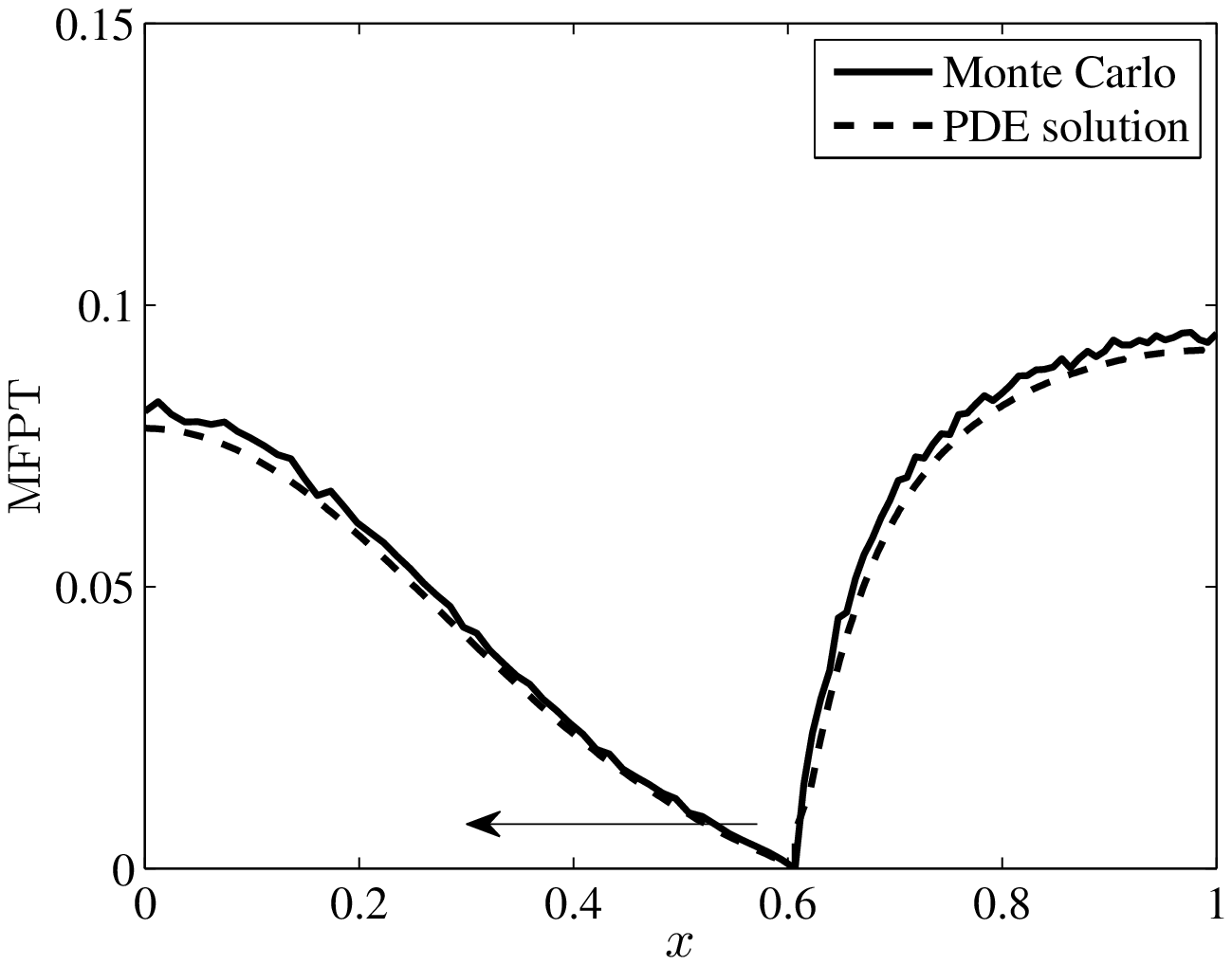}
        }   \hspace{.5cm}
    \subfigure[trap moving rightward] 
        {\label{dynslice2}
        \includegraphics[width=.3\textwidth]{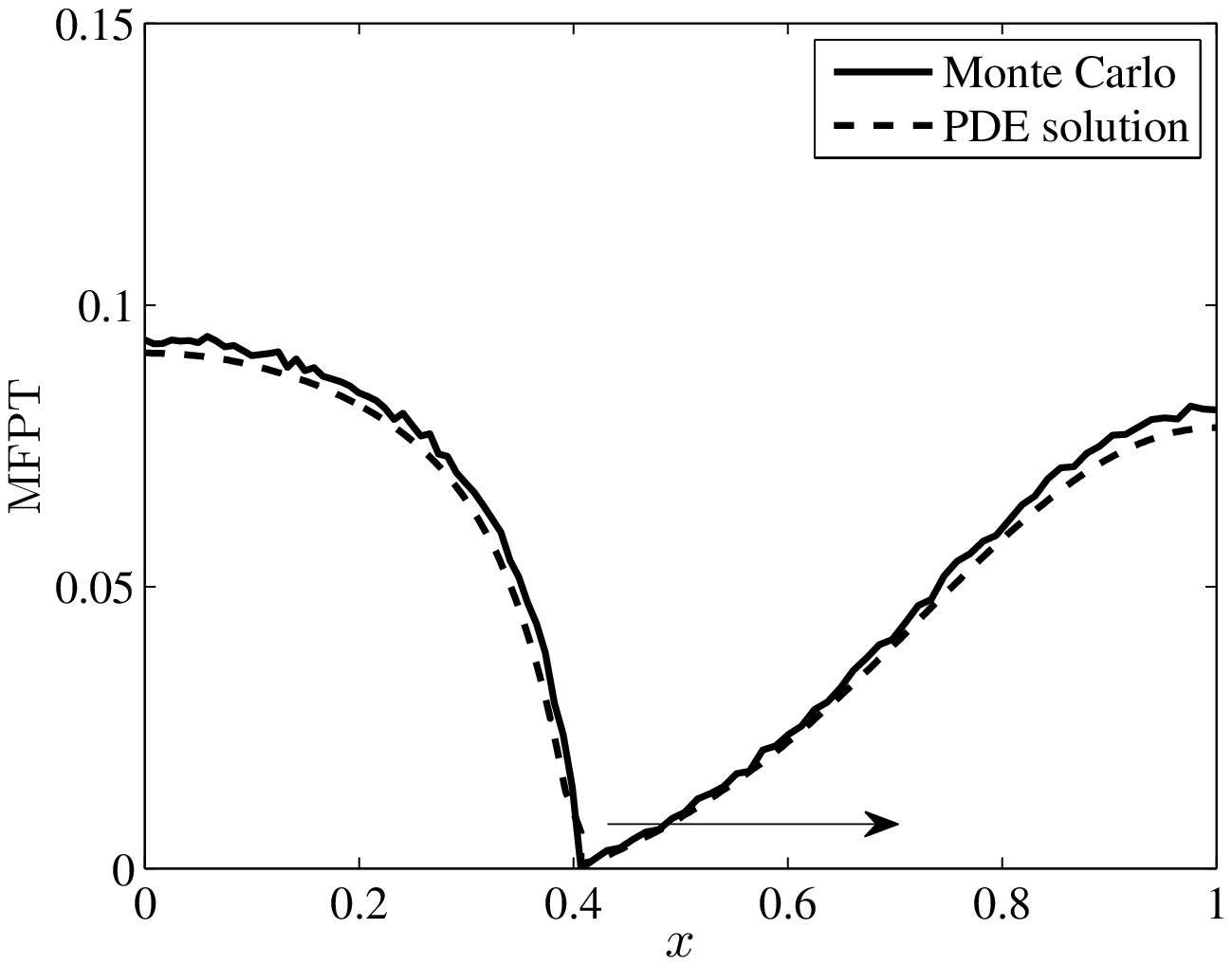}
        }
    \subfigure[surface plot of MFPT] 
        {\label{pdesurf_osc}
        \includegraphics[width=.3\textwidth]{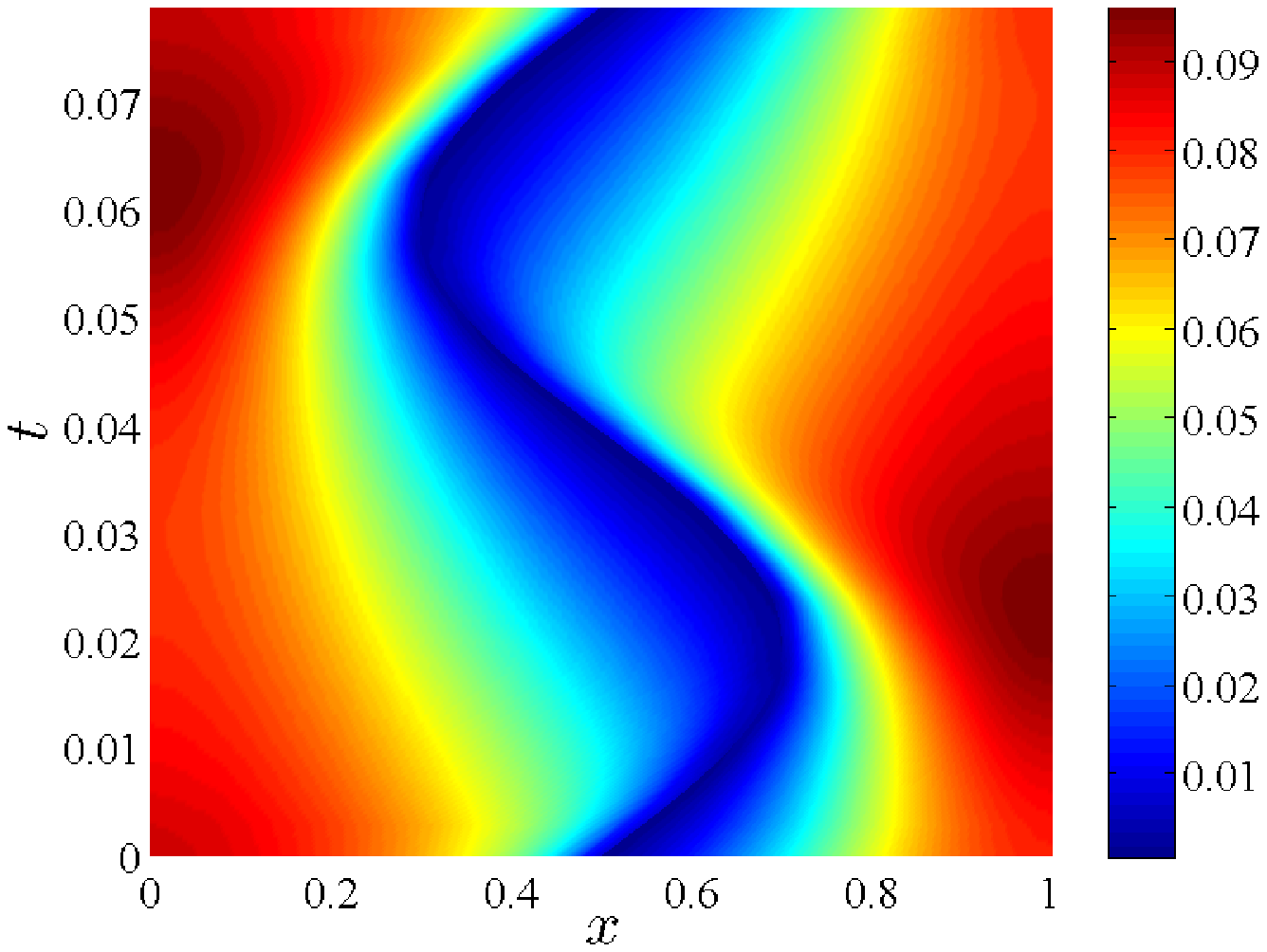}
        }}
    \caption{Comparison of MFPT as given by the PDE solution of \eqref{pbmA} with $\ve = 0.2$ and $\omega = 80$ (dashed) versus that given by Monte Carlo simulations (solid). (a) MFPT when the trap is located near $x = 0.5$ and moving leftward. (b) MFPT when the trap is located near $x = 0.5$ and moving rightward. Note that the MFPT is high (low) directly behind (in front of) the trap. (c) Space-time representation of MFPT from a numerical solution of \eqref{pbmA}.}
  \end{center}
\end{figure}\end{empty}

\subsection{Critical oscillation frequency}

As stated in the introduction, our goal is to determine the threshold frequency $\omega_c$ for which a trap oscillating about $x = 1/2$ with amplitude $\varepsilon \ll 1$ and frequency $\omega > \omega_c$ performs more optimally than a stationary trap located at $x = 1/2$. To facilitate analysis, we exploit the left-right symmetry (see Figure \ref{pdesurf_osc}) and consider only the left half of the domain $0 < x < 1/2 + \ve\sin\omega t$, $0 < t < 2\pi/\omega$. On this half-domain, the interior absorbing segment in Figure \ref{twoDschematic} acts as an absorbing Dirichlet boundary. The expression $\bar{u}$ in \eqref{ubar} for the average MFPT then becomes

\begin{equation} \label{ubarosc}
\bar{u}=\frac{\omega}{\pi}\int_{0}^{\frac{2\pi}{\omega}}\int_{0}%
^{1/2+\varepsilon\sin(\omega t)} \!u(x,t)\, dxdt \,.
\end{equation}

\noI To find $\omega_c$ in the limit of small $\ve$, we treat \eqref{pbmA} as a perturbed boundary problem and compute a three-term regular asymptotic expansion for $u$. That is, we expand $u(x,t)$ as

\begin{equation} \label{uexpand}
	u(x,t) = u_0(x) + \ve u_1(x,t) + \ve^2 u_2(x,t) \,.
\end{equation}

\noI In \eqref{uexpand}, 

\BE \label{u0}
u_0(x) = -x^2/2 + 1/8
\EE

\noI is the time-independent MFPT associated with a stationary trap with spatially averaged MFPT $\bar{u}_0 = 1/12$. With \eqref{uexpand}, $\bar{u}$ in \eqref{ubarosc} has the small-$\ve$ expansion,

\begin{equation}\label{ubarpiexpand}
	\bar{u} \sim \frac{1}{12} + \ve \frac{\omega}{\pi}\int_0^{\frac{2\pi}{\omega}}\int_0^{1/2}\! u_1(x,t) \, dxdt  + \ve^2\frac{\omega}{\pi}\left\lbrack \int_0^{\frac{2\pi}{\omega}}\int_{0}^{1/2}\!u_2(x,t)\,dxdt  + \int_0^{\frac{2\pi}{\omega}}\!u_1(1/2, t)\sin \omega t \, dt - \frac{1}{8} \right\rbrack  \,.
\end{equation}

\noI We show below that the $\mathcal{O}(\ve)$ term in \eqref{ubarpiexpand} evaluates to zero. The condition on the leading order threshold value of $\omega_c$ must then occur when the $\mO(\ve^2)$ term is also zero. 

To obtain this condition, we must calculate $u_1$ and $u_2$ in \eqref{ubarpiexpand}. The insulating condition at $x = 0$ and the periodicity condition in $t$ remain unchanged for $u_i$, $i = 0, 1, 2$. For the Dirichlet condition at $x = \xi(t)$ in \eqref{pbmAbc}, we expand for small $\ve$ and collect powers to obtain the following boundary value problems:

\bes
\BE \label{u0eq}
u_{0xx} + 1 = 0 \,; \qquad u_{0x}(0) = 0 \,, \quad u_0(1/2) = 0 \,; 
\EE
\BE \label{u1eq}
u_{1xx} + u_{1t} = 0 \,; \qquad u_{1x}(0, t) = 0 \,, \quad  u_{1}(1/2,t) = -\sin\left(\omega t\right)  u_{0x}(1/2) \,, \quad u_1(x,0) = u_1(x,2\pi/\omega) \,;
\EE
\begin{multline} \label{u2eq}
u_{2xx} + u_{2t} = 0 \,; \qquad u_{2x}(0, t) = 0 \,, u_{2}(1/2,t) = -\sin(\omega t)u_{1x} - \\ u_{0xx}(1/2)\sin^{2}(\omega t)/2 \,, \quad u_2(x,0) = u_2(x,2\pi/\omega) \,.
\end{multline}
\ees

\noI The solution for \eqref{u0eq} is given by \eqref{u0}, while solving for $u_1$ in \eqref{u1eq} yields

\begin{equation} \label{u1}
	u_{1}=e^{it\omega}\frac{\cosh\left(  \sqrt{-i\omega}\,x\right)  }{4i\cosh(\sqrt{-i\omega}/2)}-e^{-it\omega}\frac{\cosh\left(  \sqrt{i\omega}\,x\right)}{4i\cosh(\sqrt{i\omega}/2)} \,.
\end{equation}

\noI Next, \eqref{u1} with \eqref{u2eq} suggests that $u_2$ in \eqref{u2eq} has the form

\BE \label{u2}
u_{2}(x,t) = \frac{1}{4}-\frac{1}{8}\sqrt{i\omega}\tanh\left(\sqrt{i\omega}/2\right)-\frac{1}{8}\sqrt{-i\omega}\tanh\left(\sqrt{-i\omega}/2\right) + p(x)\,e^{i2\omega t} + q(x)\,e^{-i2\omega t} \,.
\EE

\noI Since the $t$-integral in \eqref{ubarpiexpand} of the oscillatory modes in \eqref{u2} evaluate to zero, we find that

\bes \label{dominicana}
\BE
\bar{u} = \frac{1}{12}+\frac{\varepsilon^{2}}{8}h(\omega) \,,
\EE
\noI where
\BE \label{h}
h(\omega) \equiv 4-\sqrt{i\omega}\tanh\left(\sqrt{i\omega}/2\right)-\sqrt{-i\omega}\tanh\left(\sqrt{-i\omega}/2\right) \,.
\EE
\ees

\noI We show below that $h(\omega)$ in \eqref{h} is monotonic decreasing and has a unique zero. We therefore conclude that the average MFPT is lower when the trap remains stationary ($\ve = 0$) than when it oscillates with small amplitude ($0 < \ve \ll 1$) when $0<\omega<\omega_{c}$, where $\omega_{c}\approx9.6017$ satisfies

\BE \label{omegac}
 h(\omega_c) = 0  \,.
\EE
 
\begin{figure}[htbp]
  \begin{center}
  \mbox{
     \subfigure[$\ve^{-2}(\bar{u} - \bar{u}_0)$ versus $\omega$]
     {\label{fig:oeps2mass}
     \includegraphics[width=.4\textwidth]{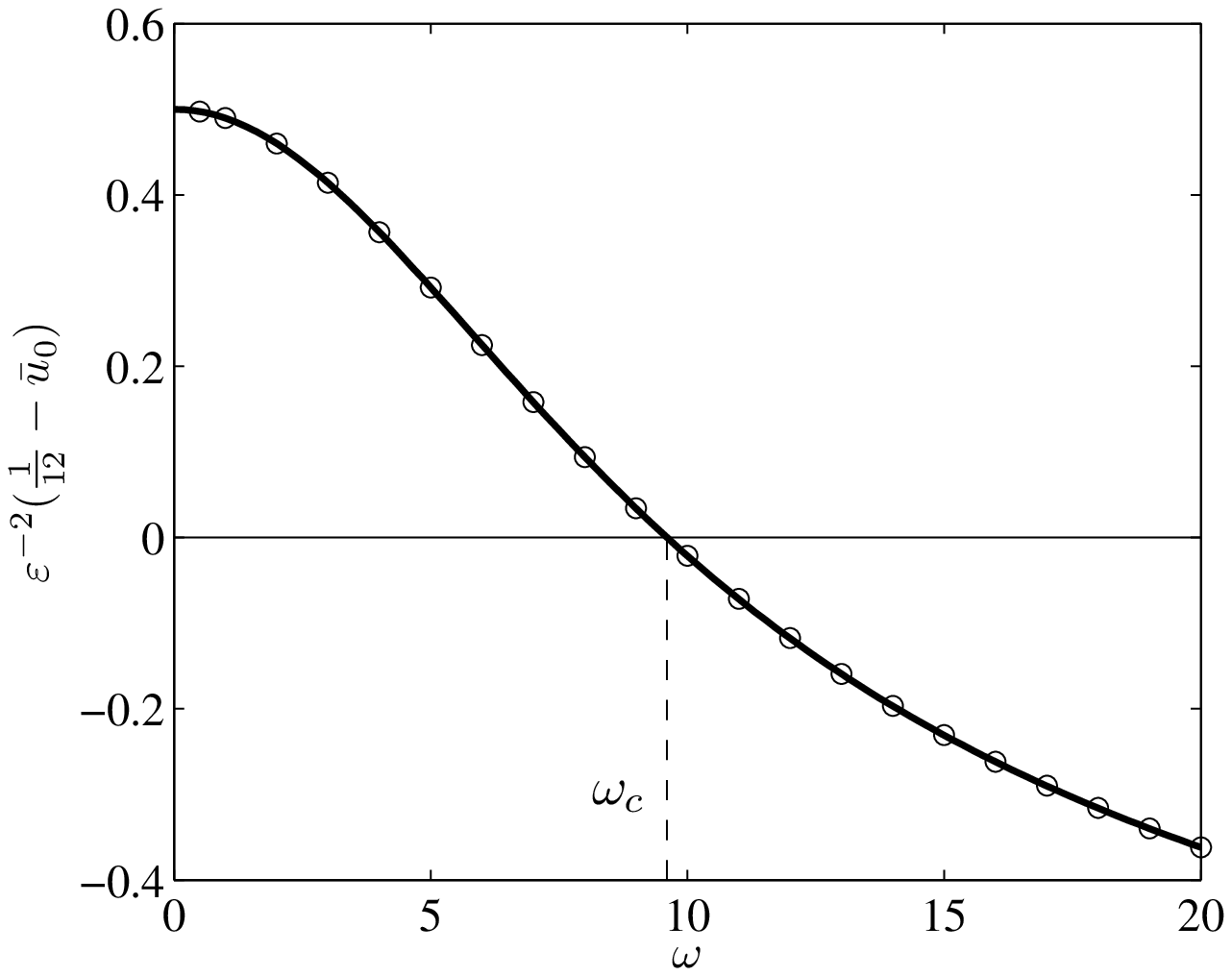}
     } \hspace{1cm}
     \subfigure[$\omega_c$ versus $\ve$]
     {\label{omega_c_O1eps}
        \includegraphics[width=.4\textwidth]{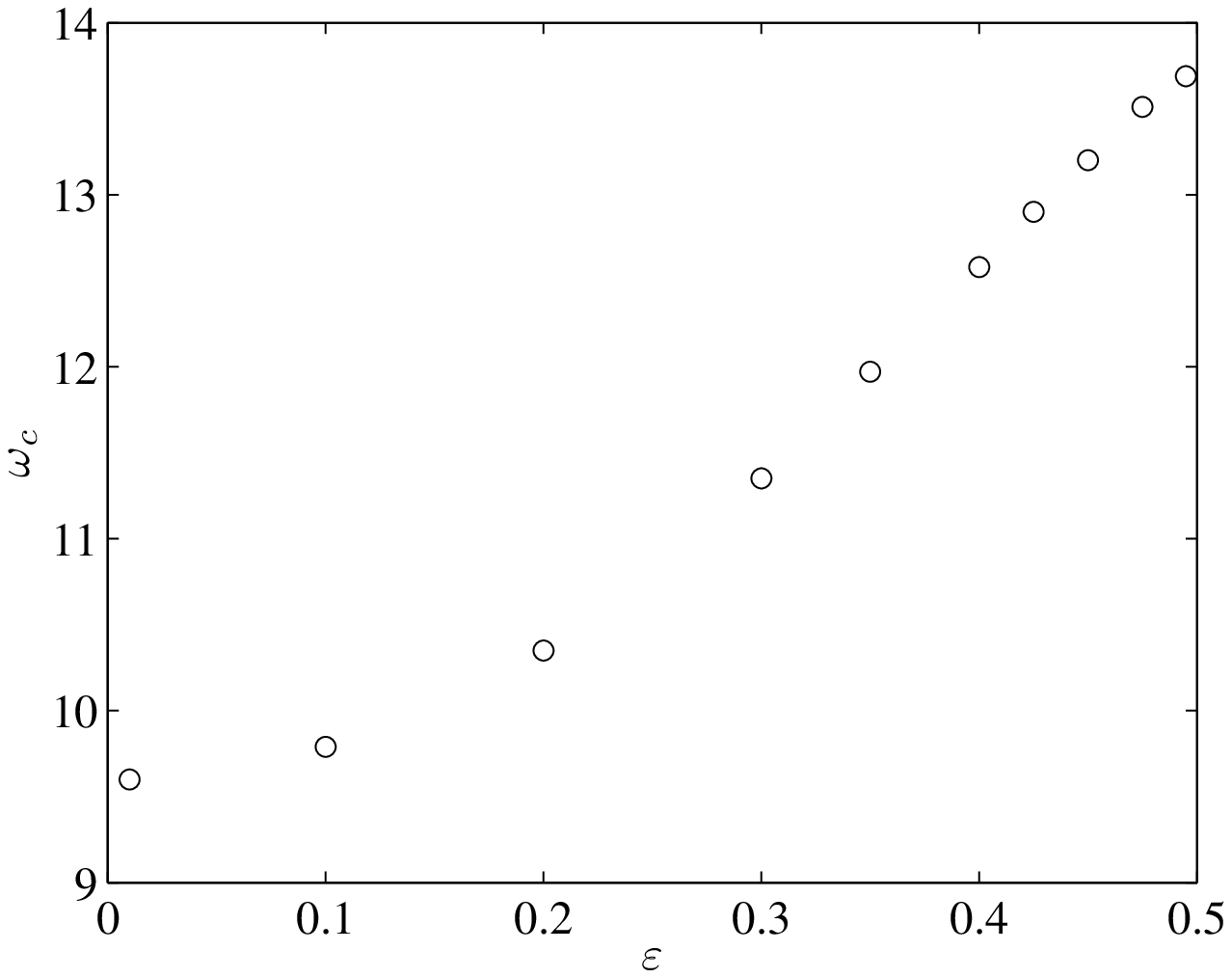}
     }}
    \caption{(a) Comparison between numerically computed values of the quantity $\ve^{-2}(\bar{u} - \bar{u}_0)$ (circles) and the asymptotic result given by \eqref{dominicana} (solid curve). Here, $\ve = 0.01$. Note the axis crossing near $\omega \approx \omega_c$. (b) Critical oscillation frequency $\omega_c$ for a range of oscillation amplitude $\ve$. Observe that for $\ve$ small, $\omega_c \sim 9.6017$.}
  \end{center}
\end{figure}

\noI Conversely, the average MFPT is lower in the presence of an oscillating trap when $\omega > \omega_c$. In Figure \ref{fig:oeps2mass}, we show a favorable comparison between numerically computed values of the quantity $\ve^{-2}(\bar{u} - 1/12)$ (circles) and the asymptotic result given by \eqref{dominicana} (solid curve). The numerical results were obtained from the FlexPDE finite element solver. We note that both plots cross the horizontal axis near $\omega \approx \omega_c$.

To show that the positive root to $h(\omega)$ exists and is unique, we find an
alternative solution to (\ref{pbmA})\ using the Fourier series in space
(instead of separation of variables in time). After some algebra, we obtain

\begin{equation}
h(\omega)=8\left\{  \frac{1}{2}-\sum_{k=0}^{\infty
}\frac{1}{\left(  2k+1\right)  ^{4}/\omega^{2}+1}\right\} \,.\label{republicana}%
\end{equation}

\noI The equivalence of (\ref{republicana}) and (\ref{dominicana}) 
can also be seen directly by using the identity

\begin{equation}
\tanh(s)=\sum_{k=0}^{\infty}\frac{2s}{s^{2}+(\frac{2k+1}{2})^{2}%
} \,.\label{tanhsum}%
\end{equation}

\noI From \eqref{republicana}, we observe that $h(\omega)$ is monotonic decreasing since each individual term in the sum is monotonic increasing in $\omega$. Next, we calculate the large $\omega$ asymptotics of the sum in \eqref{republicana} as

\BE \label{lomega}
\sum_{k=0}^{\infty}\frac{1}{\left(  2k+1\right)^{4}/\omega^{2}+1}\sim
\frac{\sqrt{\omega}}{2}\int_{0}^{\infty}\!\frac{1}{1+s^{4}}\,ds \sim\frac{\pi}%
{4}\sqrt{2\omega} \,, \quad \omega\gg 1 \,.
\EE

\noI With \eqref{republicana}, \eqref{lomega} shows that $h(\omega) \to -\infty$ as $\omega \to \infty$. With $h(0) = 4 > 0$, we thus conclude that $h(\omega)$ is a decreasing function of $\omega$, positive for small $\omega$ and negative for sufficiently large $\omega$. This proves the existence and uniqueness of a positive root of $h(\omega)$.

In \S \ref{movingsink}, we show that the critical frequency $\omega_c$ is identical to a Hopf bifurcation frequency of certain regime of the Gray-Scott model, the resulting reduced system of which is given in \eqref{pbmC}. The problem of splitting probability is considered briefly in the discussion section. We show that, in contrast to the MFPT problem, the splitting probability does not exhibit the type of behavior characterized by the existence of a critical oscillation frequency.

\setcounter{equation}{0}
\section{Brownian trap} \label{brown}

In this section, we consider the case of a Brownian trap. To derive a boundary value problem describing the MFPT, we adopt the same
approach as in \cite{giuggioli2013encounter,gabel2012can}. At each instant in time the system may be
defined by the locations $x$ and $y$ of the particle and trap, respectively.
Assuming that each undergoes an unbiased discrete random walk, the system in
state $(x,y)$ may move to one of its nearest neighbors every $\Delta t$ time
step. The two-agent random walk in one dimension can then be mapped onto a
one-agent random walk in two dimensions on a rectangular lattice with
horizontal and vertical spacings $\Delta x$ and $\Delta y$. The PDE for the
MFPT $v(x,y)$ may then be readily obtained in the same manner as in
\S \ref{sec:osc}. We write%

\begin{equation}
v(x,y)=\frac{1}{4}\left(  v(x+\Delta x,y)+v(x-\Delta x,y)+v(x,y+\Delta
y)+v(x,y-\Delta y)\right)  +\Delta t\,,\label{mfptranddisc}%
\end{equation}
which states that the MFPT associated with the state $(x,y)$ is the average of
that of the four states it may evolve to next, plus the $\Delta t$ time that
elapses between the transition. Expanding \eqref{mfptranddisc} for small
$\Delta x$ and $\Delta y$ we obtain%

\begin{equation}
\frac{\left(  \Delta x\right)  ^{2}}{2\Delta t}v_{xx}+\frac{\left(  \Delta
y\right)  ^{2}}{2\Delta t}v_{yy}+1=0\,.\label{mfptcontunscaled}%
\end{equation}

\noI Rescaling $v$ in \eqref{mfptcontunscaled} by $v=2\Delta t/\left(  \Delta x\right)  ^{2}u$ we then obtain the
boundary value \eqref{pbmB} with $a=\Delta y/\Delta x$.

\begin{figure}[ptbh]
\begin{center}
\includegraphics[width=.4\textwidth]{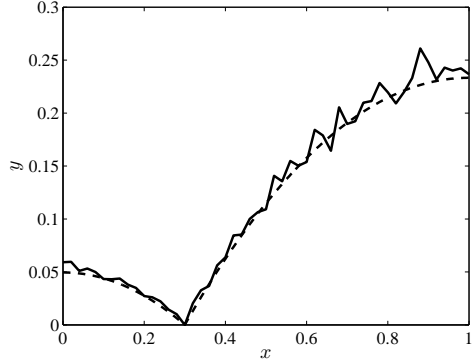}
\end{center}
\label{mfpt_simul_vs_pde}
\caption{Comparison of the numerical solution of \eqref{pbmB} with ${a^{2}%
}=0.1$ (dashed) against MFPT results from Monte Carlo simulations (solid). The starting
location of the trap is $y_{0}=0.3$.}%
\end{figure}

In Figure \ref{mfpt_simul_vs_pde}, we compare the numerical solution of
\eqref{pbmB} with ${a^{2}}=0.1$ against MFPT results from Monte Carlo
simulations. In the Monte Carlo simulations, the MFPT associated with a given
point $x$ was computed by averaging over $500$ realizations of a randomly
diffusing particle-trap pair starting from respective locations $x$ and
$y_{0}$. In Figure \ref{mfpt_simul_vs_pde}, $y_{0}=0.3$. The capture time of a
particular realization was taken to be the time elapsed before the particle
and trap locations first coincided.

The main question that we seek to address is:\ \emph{For a given speed ratio
}$a$ $\emph{and}$\emph{\ an initial trap location }$y_{0}$, \emph{is it more optimal
for a trap to move randomly or to remain stationary?} As in \S \ref{sec:osc}, we address
this question by comparing the \emph{average} MFPT over $x$. That is, for a
given $a$ and $y_0$, we compare $\bar{u}(y_{0};a)=\int_{0}^{1}\!u(x,y_{0})\,dx$ for zero and nonzero $a$. Using the
underlying symmetry $u(x,y)=u(1-y,1-x)$ to reduce the domain from a square to
a triangle $0<x<y,\ 0<y<1$, this average may then be calculated as

\begin{equation} \label{avgmfptrand}
\bar{u}(y_{0};{a})=\int_{y_{0}}^{1}\!u(x,y_{0})\,dx+\int_{1-y_{0}}%
^{1}\!u(x,1-y_{0})\,dx\,.
\end{equation}

By computing $\bar{u}$ numerically, we find that there is a critical value $a_{c}\approx0.53$ such that $\bar{u}\left(  1/2,a_{c}\right)  =\bar
{u}\left(  1/2,0\right)  $ and moreover that $\bar{u}\left(  y_{0},a\right)
<\bar{u}\left(  y_{0},0\right)  $ for \emph{any }$y_{0}$, as long as
$a>a_{c}$. That is, regardless of initial trap location, a mobile trap
is more optimal (in the sense of reducing $\bar{u}(y_0;a)$) than a stationary trap as long as the trap moves quickly enough. When $a<a_{c}$, there exists $y_{c}(a)$ such that $\bar{u}(y_{0};a)>\bar{u}(y_{0};0)$ as long as $y_{0}\in\left(y_{c},1-y_{c}\right)$. Moreover, we find analytically that $y_{c}(a)\rightarrow0$ as $a\rightarrow0$. We thus conclude that a very slow trap performs worse than a stationary trap except when when its initial
location is very close to the boundary. In this limit of a very slow trap, we determine the asymptotic formula for $y_c$

\begin{equation}\label{yc} 
	y_{c}\sim\frac{2}{\pi}a\log\left\lbrack\frac{48}{\pi^3 a} \right\rbrack \,,\qquad a\ll1 \,. 
\end{equation}

\begin{figure}[htbp]
  \begin{center}
  	\mbox{
  	\subfigure[asymptotic versus numerical $y_c$]
        {\label{fig:y0c}
        \includegraphics[width=.4\textwidth]{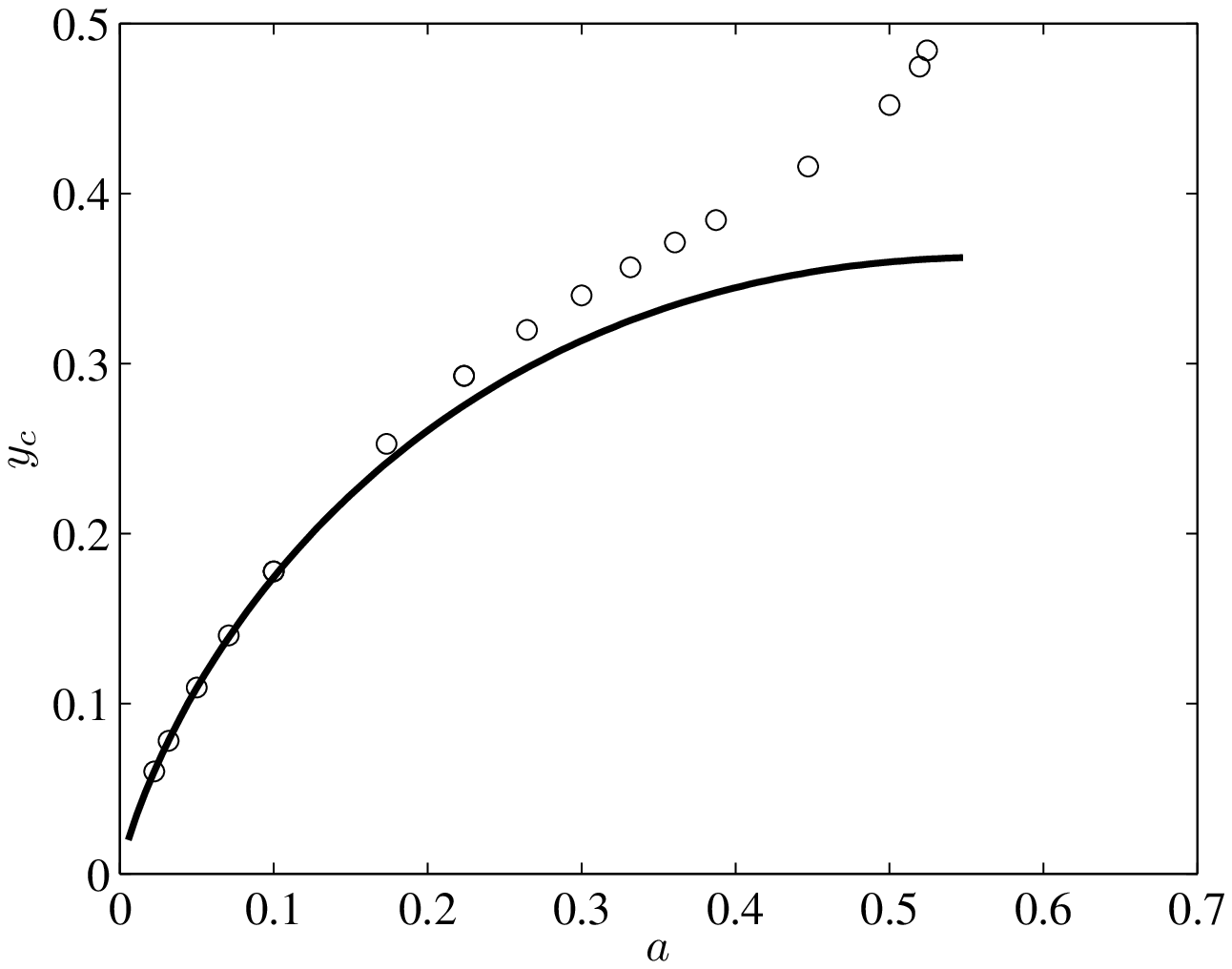}
        }   \hspace{1cm}
    \subfigure[higher order calculation of $y_c$]
    		{\label{optbound_usethis_moreaccurate}
        \includegraphics[width=.4\textwidth]{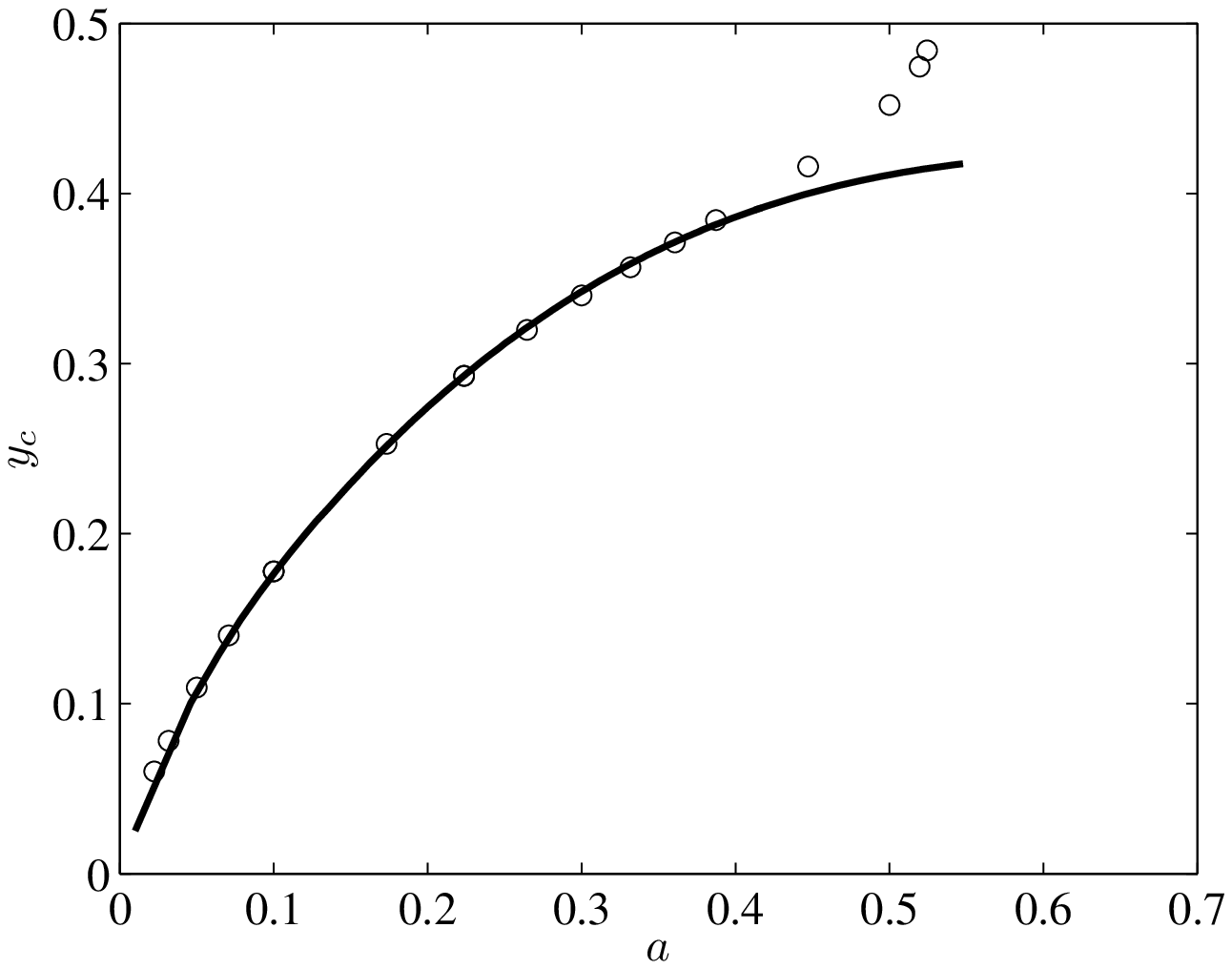}
        }}
    \caption{(a) Comparison of the asymptotic solution (solid curve) for $y_c$ as given by \eqref{yc} and the numerical result (circles) obtained by solving the full PDE \eqref{pbmB}. For trap starting location $y_0 < y_c$ ($y_0 > y_c$), the mobile (stationary) trap is more optimal on average. For $a \gtrsim 0.53$, the mobile trap is always more optimal. (b) Comparison of the asymptotic solution (solid curve) for $y_c$ as given by a numerical solution of \eqref{y0ceq}. Compared to (a), we observe a greater range of agreement with the numerical results.}
  \end{center}
\end{figure}

\noI These results are summarized in Figure \ref{fig:y0c}. 

The derivation of (\ref{yc})\ requires the analysis of the contribution of the
boundary layer of \eqref{pbmB} near $y=\mO(a)$. In the outer region $y\gg
\mO(a)$, we write $u \sim u^o$ with $u^o \sim u_{0}+{a^{2}}u_{1}+O(a^{4})$. From \eqref{pbmB},
both $u_{0}$ and $u_{1}$ satisfy the boundary value problem%

\begin{equation}
u_{xx}+1=0\,, \qquad 0<y<1 \,, \quad y<x<1\,, \quad u_{x}(1,y)=0\,,\quad
u(y,y)=0\,.\label{uo0eq}
\end{equation}

\noI Solving \eqref{uo0eq}, we obtain for the outer solution

\begin{equation}
u^o \sim (1+a^2)\left\lbrack \frac{y^2}{2} - \frac{x^2}{2} + x-y \right\rbrack \,. \label{outersol}%
\end{equation}

\noI We note that \eqref{outersol} does not satisfy the no-flux boundary
condition in \eqref{pbmBbc} on the line $y=0$, which will be satisfied by the
boundary layer. In this inner region where $y = \mO(a)$, we let $u \sim U$ where we expand $U$ as

\BE \label{Uinnerexpand}
 U(x,\eta)=U_{0}+aU_{1}+a^{2}U_{2}\ldots \,; \qquad y = a\eta \,.
\EE

\noI Next, expanding the Dirichlet boundary condition $U(a\eta,\eta) = 0$ along the diagonal $x=y$, we obtain the following boundary value problems,

\bes \label{Ueq}
\BE \label{U0eq}
U_{0xx}+U_{0\eta\eta}+1 = 0 \,; \qquad U_{0}(0,\eta)=0 \,, \quad U_{0x}(1, \eta) = 0, \quad U_{0\eta}(x,0)=0\,;
\EE
\BE \label{U1eq}
U_{1xx}+U_{1\eta\eta}=0 \,; \qquad U_{1}(0,\eta)=-\eta U_{0x}(0,\eta)\,, \quad U_{1x}(1, \eta) = 0, \quad U_{1\eta}(x,0)=0\,;
\EE
\BE \label{U2eq}
U_{2xx}+U_{2\eta\eta}=0 \,; \qquad U_{2}(0,\eta)=-\frac{\eta^{2}}{2}U_{0xx}(0,\eta)-\eta U_{1x}(0,\eta) \,, \\ \quad U_{2x}(1, \eta) = 0, \quad U_{2\eta}(x,0)=0\,.
\EE
\ees

\noI  To determine the large-$\eta$ behaviour for \eqref{Ueq}, we write \eqref{outersol} in terms of inner variables as

\BE \label{etaasymp}
U\sim x-\frac{x^{2}}{2}-a\eta+a^{2}\left(  x-\frac{x^{2}}{2}+\frac{\eta^{2}}{2}\right) \,.
\EE

\noI Matching powers of $a$ in \eqref{Uinnerexpand} and \eqref{etaasymp} suggests that we write

\BE \label{subs}
U_{0}  =x-\frac{x^{2}}{2}+V_{0} \,, \qquad U_{1}  =-\eta+V_{1}(x,\eta) \,, \qquad U_{2}  =x-\frac{x^{2}}{2}+\frac{\eta^{2}}{2}+V_{2}(x,\eta) \,,
\EE

\noI where $V_{i}\rightarrow0$ as $\eta\rightarrow\infty$. Substituting \eqref{subs} into \eqref{Ueq}, we find that $V_{0}=0$ and

\BE \label{V1}
V_{1}(x,\eta)=-\sum_{n=0}^{\infty}{\frac{2}{\lambda_{n}^{2}}e^{-\lambda_{n}\eta}\sin\lambda_{n}x} \,; \qquad \lambda_{n}\equiv\frac{(2n+1)\,\pi}{2} \,.
\EE

\noI We will see later that to determine 
the leading order value for $y_c$, it is not necessary to compute $V_2$.
It is instead computed
in Appendix \ref{ycnextorder}, where we determine $y_c$ to higher
accuracy.

Finally, we construct a uniform solution using the Van Dyke matching principle
by writing $u_{unif}=u^o+U-u_{c}$, where $u_{c}$ is the common part obtained by expanding the outer solution in inner variables or, equivalently, the inner solution in outer variables. Up to $\mathcal{O}(a^2)$, we then obtain

\bes \label{ufinal}
\begin{equation}
u_{unif}=\frac{y^{2}}{2}-\frac{x^{2}}{2}+x-y+ u_m(x,y;a) \,, \label{uunif}%
\end{equation}
\noI where
\BE \label{um}
  u_m(x,y;a) = aV_{1}(x,y/a){+a}^{2}\left(\frac{y^{2}}{2}-\frac{x^{2}}{2}+x-y + V_2(x,y/a)\right) \,.
\EE
\ees

\noI In \eqref{ufinal}, $u_m(x,y;a)$ accounts for the effect of trap mobility on the MFPT, while $V_1(x,\eta)$ is given by \eqref{V1}. The value of $y_0$ at which $u_m(x,y_0;a)$ has no effect on $\bar{u}(y_0;a)$ is precisely the critical value that determines where the mobile trap becomes more optimal in comparison to the stationary trap. Substituting \eqref{ufinal} into \eqref{avgmfptrand}, we obtain that the critical value $y_0$ satisfies

\BE \label{udiff}
-\frac{2}{\lambda_{0}^{3}}\,e^{-\lambda_{0}\frac{y_{0}}{a}}+\frac{1}{3}a = 0\,.
\EE

\noI In \eqref{udiff}, we have used the fact that under the assumption $y_0 \gg a$, $V_{1}\left(  x,y/a\right)  \sim -(2/\lambda_0^2)\,e^{-\lambda _{0}y/a}\sin\lambda_{0}x$ and that $V_{2}(x,y/a)\ll \mathcal{O}(1)$. The solution of \eqref{udiff} for $y_0$ is given by $y_c$ in \eqref{yc}, with $\bar{u}(y;a)<\bar{u}(y;0)$ if $y<y_{c}$. That is, the mobile trap is more optimal only if its starting location $y$ is sufficiently close to one of the boundaries. By \eqref{yc}, we find that the assumption $y_0 \gg a$ is indeed self consistent. This completes the derivation of the formula (\ref{yc}).

In Figure \ref{fig:y0c}, we plot the asymptotic solution (solid curve) for
$y_{c}(a)$ in \eqref{yc} and compare it to
the numerical result (circles) obtained by solving the full PDE
\eqref{pbmB}. We remark that the agreement is excellent even for moderately
large values of ${a}$. In Appendix \ref{ycnextorder}, we compute $y_c(a)$ accurate for a larger range of $a$ by calculating $V_2$. Note that, due to the reflection symmetry about $y = 1/2$, the mobile trap in this case is always more optimal than the stationary trap as long as $a > a_c \sim 0.53$. While we are only able to determine this value numerically, its existence may be ascertained as follows. First, we calculate from \eqref{outersol} and \eqref{avgmfptrand} that 

\BE \label{ua}
\bar{u}(y_0;a) = (1+a^2)\left(\frac{1}{3}-y_{0}+y_{0}^{2}\right) \,, \qquad a \ll 1 \,.
\EE

\noI In calculating $\bar{u}(y_0;a)$ in \eqref{ua}, we have let $a \to 0$ while keeping $y_0$ constant so that the boundary layer need not be considered. Since $a^2 > 0$, we observe from \eqref{ua} that for \emph{any} given initial trap location $y_0$, trap motion leads to a higher average MFPT if the speed is sufficiently small. Further, we note that $\bar{u}(y_0;0) \geq 1/12$. However, in the other limit $a \to \infty$, we calculate (by letting $u \to u/a^2$ in \eqref{pbmB} and noting that this rescaling simply reverses the roles of $x$ and $y$ in the analysis) that 

\begin{equation} \label{la}
\bar{u}\left(y_{0};a\right)  \sim\frac{1}{6a^2} + \mO(a^{-4}) \,, \qquad a\gg1 \,.
\end{equation}

\noI In the limit of high trap mobility, \eqref{la} shows that the average MFPT approaches zero asymptotically in $a$. Therefore, $\bar{u}\left(y_{0};a\right) < \bar{u}(y_0;0)$ as $a \to \infty$. There then must exist at least one intermediate value $a_c$ for which $\bar{u}(y_0;a_c) = \bar{u}(y_0;0)$.

%
%
%

\setcounter{equation}{0}
\section{Moving sink problem} \label{movingsink}

We now compute the Hopf bifurcation value for $\beta$ for Problem C. The derivation of the reduced dynamics \eqref{pbmC} from the Gray-Scott model is given in Appendix \ref{gsderiv}. We show that the Hopf bifurcation frequency is given precisely by \eqref{omegac} with $h(\omega)$ defined in \eqref{h}.

The steady state is given by $\xi(t)=1/2$ and $u(x,t)=u_{0}(x)+C$ where $C$ is
any constant and%

\begin{equation}
u_{0}(x)=\left\{
\begin{array}
[c]{c}%
\frac{\left(  x-1\right)^{2}}{2} ,, \quad x>\frac{1}{2}\\
-\frac{x^{2}}{2} \,, \quad  x<\frac{1}{2}%
\end{array}
\right. \,.
\end{equation}
Linearizing around this steady state we let%

\BEU
\xi(x)=\frac{1}{2}+\eta e^{\lambda t}\,; \quad u\left(x,t\right)=u_{0}(x)+\phi(x)\, e^{\lambda t}\,, \qquad \eta\,,\phi\ll 1 \,,
\EEU

\noI to obtain the eigenvalue problem 

\bes\label{lamimplicit}%
\BE \label{philine1}
\lambda\phi =\phi_{xx} - \delta^{\prime}\left(\frac{1}{2}-x\right) \eta \,,
\EE
\BE
\phi^{\prime}(0)  =\phi^{\prime}(1)=0 \,,
\EE
\BE \label{linearhole}
\lambda\eta =-2\beta\eta+\beta\left[  \phi_{x}^{+}\left(  \frac{1}%
{2}\right)  +\phi_{x}^{-}\left(  \frac{1}{2}\right)  \right] \,.
\EE
\ees 

\noI The equation in (\ref{philine1}) is equivalent to removing $\delta^{\prime}%
$\ and replacing it by the jump conditions $\phi^{+}\left(  1/2\right)
-\phi^{-}\left(  1/2\right)  =\eta$, and $\phi_{x}^{+}\left(  1/2\right)
=\phi_{x}^{-}(1/2)$. The solution is then given by
\[
\phi(x)=\frac{\eta}{2\cosh\left(  \sqrt{\lambda}/2\right)  }\left\{
\begin{array}
[c]{c}%
-\cosh\left(  \sqrt{\lambda}(x-1)\right)  ,\ \ \ x>\frac{1}{2}\\
\cosh\left(  \sqrt{\lambda}x\right)  ,\ \ x<\frac{1}{2}%
\end{array}
\right.  \,.
\]

\noI Substituting into \eqref{linearhole} yields%

\BE \label{lam}
\lambda=-2\beta+\beta\sqrt{\lambda}\tanh\left(  \sqrt{\lambda}/2\right) \,.
\EE

To find the Hopf bifurcation frequency $\omega_H$, we set $\lambda=i\omega_H$ in \eqref{lam} and calculate the Hopf bifurcation threshold 

\bes
\BE \label{beta}
\beta_H=\frac{i\omega_H}{-2+\sqrt{i\omega_H}\tanh\left(  \sqrt{i\omega_H}/2\right)} \,.
\EE
\noI The corresponding frequency is determined by imposing that $\beta$ must be real. That is, setting the imaginary part of the right-hand side of \eqref{beta} to zero, we have
\BE \label{omega}
h(\omega_H) \equiv
4-\sqrt{i\omega_H}\tanh\left(\sqrt{i\omega_H}/2\right)-\sqrt{-i\omega_H}\tanh\left(\sqrt{-i\omega_H}/2\right) =0\,,
\EE
\ees 

\noI where $h(\omega)$ is the same function as that defined in \eqref{h}. By \eqref{omega}, we thus find that the equation for the Hopf bifurcation frequency $\omega_H$ is identical to that obtained for $\omega_{c}$ in
(\ref{omegac}). In particular, as shown above by \eqref{republicana}-\eqref{lomega}, the positive solution for $\omega_H$ exists and is unique. This also proves the uniqueness of the Hopf bifurcation threshold in a particular regime of the original Gray-Scott model.

\setcounter{equation}{0}
\section{Discussion}

It is often assumed that trap motion improves capture time (\cite{blythe2003survival, moreau2003pascal, moreau2004lattice}). Using two simple
examples (oscillatory and random trap motion), we have shown that when finite domains are considered, this may
not always the case: a mobile trap can lead to a slower capture time if its mobility is sufficiently low. In particular, for a trap located at the center of the domain, we find that undergoing random motion impairs its capture time if the motion is not sufficiently fast (more than $0.53$ times the speed of the particle). For a trap under prescribed motion oscillating with small amplitude about the center of the domain, the effect of motion on the average MFPT also depends on the mobility of the trap: a mobile trap leads to a higher average MFPT when its oscillation frequency $\omega$ is less than some critical frequency $\omega_c \approx 9.6017$. When the oscillation amplitude is not small, the dependence of $\omega_c$ on the amplitude is shown Figure \ref{omega_c_O1eps}, where the individual points were computed from full numerical solutions of \eqref{pbmB}. The increasing behavior of $\omega_c$ with $\ve$ reflects the fact that the more the trap deviates from its optimal location at the center of the domain, the faster it needs to move in order recover to the center sufficiently quickly. Observe that for $\ve$ small, $\omega_c \sim 9.6017$.

We also showed a surprising connection between the MFPT problem with oscillating interior trap and the frequency of oscillations in spike position for the Gray-Scott model. In particular, we showed that the critical and bifurcation frequency in the respective problems are identical. By showing this equivalence, we were able to prove the existence and uniqueness of a Hopf bifurcation in a particular limit of the Gray-Scott model.

The existence of a critical oscillation frequency in Problem A might suggest a similar type of behavior when considering the problem of splitting probability in the presence of a stationary and oscillatory trap. However, we show briefly here that in the limit of small oscillation amplitude, the random walker is always more likely to become trapped by the mobile trap instead of the stationary one. This problem
consists of a trap at the right boundary whose position is given by
$\xi\left( t\right)  =\frac{1}{2}+\varepsilon\sin(\omega t)$, $\ve \ll 1$, and a stationary trap at $x=0$ on the left. By analogy to the derivation for the MFPT problem \eqref{pbmA}, and that given in \cite{redner2001guide}, we obtain the ODE for the (rescaled) splitting probability

\bes \label{spliteq}
\BE \label{splitde}
u_{t} + u_{xx}=0 \,;
\EE
\BE \label{splitbc}
u(0,t)=1 \,, \quad u\left(1/2+\varepsilon\sin\left(\omega t\right), t\right) = 0 \,, \quad u(x,0) = u(x, 2\pi/\omega) \,.
\EE
\ees

\noI In \eqref{spliteq}, $u(x,t)$ gives the probability that a random walker starting at location $x$ at time $t < 2\pi/\omega$ gets captured by the stationary trap located at $x = 0$. We find that the average of this probability over one period of trap oscillation, given by

\BE \label{ubarsplit}
\bar{u}=\frac{\omega}{\pi}\int_{0}^{\frac{2\pi}{\omega}}\int_{0}%
^{1/2+\varepsilon\sin(t\omega)} \!u(x,t) \,dxdt \,,
\EE

\noI is less than $1/2$ for any $\omega$. Indeed, using a computation
similar to that performed in \S \ref{sec:osc}, we find that
\begin{equation}
\bar{u}=\frac{1}{2}+\varepsilon^{2}\left\{  1-\sqrt{i\omega}\coth\left(
\sqrt{i\omega}/2\right)  -\sqrt{-i\omega}\coth\left(  \sqrt{-i\omega
}/2\right)  \right\}\,, \qquad \varepsilon\ll1 \,.\label{dellcom}%
\end{equation}

\noI It is immediate from the identity \eqref{tanhsum} that the term in \eqref{dellcom} proportional to $\ve$, which accounts for the small amplitude oscillation of the right-hand trap, is negative for all $\omega$. We thus conclude that, in the limit of small amplitude oscillations, the mobile trap is more likely to capture the random walker than the stationary trap on the opposite side, regardless of $\omega$ (unlike
Problem A, which exhibited a threshold behaviour in $\omega$). Numerical solutions of \eqref{spliteq} with $\ve = \mO(1)$ suggest that the same conclusion also holds for $\mO(1)$ amplitude oscillations.

An interesting problem would be to see if the equivalence between Problems A and C carries over to two or more dimensions. The derivation of \eqref{pbmA} for the MFPT extends easily to higher dimensions, and thus may also be employed to formulate PDE's for the MFPT in the presence of general trap motion and domain geometry. This allows for a full investigation of MFPT problems with mobile traps by way of either numerical computations or asymptotic analysis. Such a study has not yet been carried out.

Another interesting problem would be to replace the sinusoidal trap motion in Problem A by a general periodic $f(t)$ with the same period and amplitude. This analysis could be done by writing $f(t)$ in terms of its Fourier series. An optimization problem may be then be posed by asking what $f(t)$ minimizes average MFPT, while penalizing the average square velocity of the trap. In practice, this type of problem lends to situations where reduced capture times must be weighed against higher energy expenditure, for example, in predator search strategy.

\section*{Acknowledgments}

J. C. Tzou was supported by an AARMS Postdoctoral Fellowship. T. Kolokolnikov
is supported by NSERC discovery and NSERC accelerator grants.

\appendix

\setcounter{equation}{0}
\section{Derivation of Problem C} \label{gsderiv}

We begin with the Gray-Scott model as scaled in \cite{chen2009oscillatory}:

\bes \label{GS}
\BE 
v_{t}=\varepsilon^{2}v_{xx}-v+Auv^{2} \,, \qquad v_x(0,t) = v_x(1,t) = 0 \,,
\EE
\BE
	 \tau u_{t} =Du_{xx}+1-u-\frac{uv^{2}}{\varepsilon} \,, \qquad u_x(0,t) = u_x(1,t) = 0 \,,
\EE
\ees

\noI supplemented by appropriate initial conditions. In the limit $\ve \to 0$, it is shown that \eqref{GS} admits a single spike solution whose dynamics are given by the reduced system

\begin{equation} \label{gsdyn}
u_{t}=Du_{xx}+1-u-\frac{6}{A^{2}u_{0}}\,\delta\left(  x-\xi\right) \,; \qquad  \xi_{t}=\frac{\tau\varepsilon^{2}}{u_{0}}\left(u_{x}(\xi^+(t),t) +u_{x}(\xi^-(t),t) \right) \,,
\end{equation}

\noI where $u_0 = u(\xi(t),t)$. In \eqref{gsdyn}, $\xi(t)$ represents the location of the center of the spike. We assume that $A\gg1$ in \eqref{gsdyn} and make a change of variables

\BE \label{uhat}
u=1+\frac{\hat{u}}{A^{2}} \,.
\EE

\noI With \eqref{uhat} in \eqref{gsdyn}, the leading order terms yield

\begin{equation} \label{uhateq}
\hat{u}_{t}=D\hat{u}_{xx}-\hat{u}-6\delta\left(  x-\xi\right) \,; \qquad \xi_{t}=\frac{\tau\varepsilon^{2}}{A^{2}}\left(  \hat{u}_{x}^{+}%
+\hat{u}_{x}^{-}\right)  \,.
\end{equation}

\noI Next, we suppose that $D\gg1$ and make the final change of variables in \eqref{uhateq}

\[
\hat{u}=6\left(-1+\frac{\tilde{u}}{D}\right)\,, \qquad 
t=\hat{t}/D\,,\qquad
\beta=\frac{6\tau\varepsilon^{2}}{A^{2}D^2} \,.
\]

\noI The resulting leading-order expression becomes precisely
\eqref{pbmC} upon dropping the hats.

\setcounter{equation}{0}

\section{Next order term for computation of $y_c$} \label{ycnextorder}

A more accurate formula for $y_{c}$ can be obtained by keeping all terms from
(\ref{uunif}) in (\ref{udiff}). In particular this requires the full
computation of $V_{2}$ in \eqref{subs}. Substituting \eqref{subs} for $U_2$ into \eqref{U2eq}, we obtain the equation for $V_2$

\begin{equation} \label{vi2eq}
	V_{2xx} + V_{2\eta\eta} = 0 \,; \qquad V_2(0,\eta) = f(\eta) \,, \quad V_{2x}(1,\eta) = 0 \,, \quad V_{2\eta}(x,0) = 0 \,, \quad V_2 \to 0 \enspace \mbox{as} \enspace \eta \to \infty \,.
\end{equation}

\noI In \eqref{vi2eq}, $f(\eta)$ is defined as

\BE\label{feta}
f(\eta) \equiv \eta \sum_{n=0}^\infty{\frac{2}{\lambda_n}e^{-\lambda_n \eta}} \,.
\EE

\noI To solve \eqref{vi2eq}, we use a Fourier cosine transform and its inverse defined by

\begin{equation*}
	\hat{G}(x,\omega) = 4 \int_0^\infty\! g(x,\eta) \cos (2\pi\omega\eta) \, d\eta \,, \qquad g(x,\eta) = \int_0^\infty \! \hat{G}(x,\omega) \cos(2\pi\omega\eta) \, d\omega \,.
\end{equation*}

\noI Proceeding, we calculate that

\begin{equation} \label{V2sol}
V_2 = \int_0^\infty \! \frac{\hat{F}(\omega)}{\cosh(2\pi\omega)}\cosh(2\pi\omega(x-1))\cos(2\pi\omega\eta) \, d\omega \,,
\EE

\noI where $\hat{F}(\omega)$ is the Fourier cosine transform of $f(\eta)$ in \eqref{feta}. The resulting expression for $u_m(x,y;a)$, accurate to $\mO(a^2)$, is then given by \eqref{um} with $V_2(x,y/a)$ given by \eqref{V2sol}. The condition that $u_m(x,y;a)$ integrates to $0$ in \eqref{avgmfptrand} then yields the condition for $y_c$,

\begin{multline} \label{y0ceq}
	-\sum_{n=0}^\infty{\frac{2}{\lambda_n^3}e^{-\lambda_n y_c/a} \cos(\lambda_n y_c)} + a\left\lbrack (y_c)^2 - y_c + \frac{1}{3}\right\rbrack + \\ + a \int_0^\infty \! \frac{\hat{F}(\omega)}{2\pi\omega\cosh(2\pi\omega)}\sinh(2\pi\omega(x-1))\cos(2\pi\omega y_c/a) = 0 \,.
\end{multline}

\noI Solving \eqref{y0ceq} numerically for $y_c$, we obtain the solid curve in Figure \ref{optbound_usethis_moreaccurate}. Note that, in comparison to Figure \ref{fig:y0c}, which contains the leading order expression for $y_c$ in \eqref{yc}, we observe a larger range of agreement between the asymptotic and numerical (open circles) results.

%

\bibliographystyle{elsart}
\bibliography{trap7}
\end{document}